\documentclass[prd,nofootinbib,showpacs,preprint]{revtex4}
\usepackage{amsmath}
\usepackage{graphicx}
\usepackage{epsfig}
\graphicspath{{Figs/}}
\usepackage{dcolumn}
\usepackage{bm}
\usepackage{amssymb}
\usepackage{float}
\usepackage[usenames,dvipsnames]{color}
\usepackage{slashed}
\usepackage[dvipdfm,colorlinks,citecolor=blue]{hyperref}
\usepackage{booktabs}
\usepackage{multirow}
\newcommand{\mathsym}[1]{{}}
\newcommand{\unicode}[1]{{}}

\begin{document}
\title{Relating Lepton Mixing Angles with Lepton Mass Hierarchy}

\author{Debasish Borah \footnote{Present Address: Department of Physics, Indian Institute of Technology Guwahati, Assam-781039, India}}
\email{dborah@iitg.ernet.in}

\affiliation{Department of Physics, Tezpur University, Tezpur-784028, India}


\begin{abstract}
We revisit the possibility of relating lepton mixing angles with lepton mass hierarchies in a model-independent way. Guided by the existence of such relations in the quark sector, we first consider all the mixing angles, both in charged lepton and neutrino sectors to be related to the respective mass ratios. This allows us to calculate the leptonic mixing angles observed in neutrino oscillations as functions of the lightest neutrino mass. We show that for both normal and inverted hierarchical neutrino masses, this scenario does not give rise to correct leptonic mixing angles. We then show that correct leptonic mixing angles can be generated with normal hierarchical neutrino masses if the relation between mixing angle and mass ratio is restricted to 1-2 and 1-3 mixing in both charged lepton and neutrino sectors leaving the 2-3 mixing angles as free parameters. We then restrict the lightest neutrino mass as well as the difference between 2-3 mixing angles in charged lepton and neutrino sectors from the requirement of producing correct leptonic mixing angles. We constrain the lightest neutrino mass to be around $0.002$ eV and leptonic Dirac CP phase $\delta_{CP}$ such that $\sin^2{\delta_{CP}}\sim(0.35-0.50)$. We also construct the leptonic mass matrices in terms of $2-3$ mixing angles and lightest neutrino mass and briefly comment on the possibility of realizing texture zeros in the neutrino mass matrix. 

\end{abstract}
\pacs{14.60.Pq, 11.10.Gh, 11.10.Hi}
\maketitle
\section{Introduction}
Understanding the origin of flavor has been a major goal of particle physics research for last few decades. The observed patterns of quark as well as lepton mixing and the fact that they are very different from each other remain very puzzling till today. Apart from the mixing, the hierarchies in the masses of fermions in the standard model (SM) are also making the flavor problem even more puzzling. For example, there is a hierarchy of six orders of magnitude between the lightest charged lepton and the heaviest quark. The observed hierarchy between up type quarks, down type quarks and the charged leptons can be summarized as
\begin{equation}
m_u : m_c : m_t \approx 10^{-5} : 10^{-3} : 1
\nonumber
\end{equation}
\begin{equation}
m_d : m_s : m_b \approx 10^{-3} : 10^{-2} : 1
\nonumber
\end{equation}
\begin{equation}
m_e : m_{\mu} : m_{\tau} \approx 10^{-4} : 10^{-2} : 1
\end{equation}
The mixing of quarks in weak interactions is described by the Cabibbo-Kobayashi-Maskawa (CKM) matrix \cite{CKM1, CKM10}. Using the Wolfenstein parameterization \cite{wolfen1}, the quark mixing angles are given by
$$ s^{\text{CKM}}_{12} = \lambda, \;\; s^{\text{CKM}}_{23} = A\lambda^2, \;\; s^{\text{CKM}}_{13} = \lvert A\lambda^3(\rho+i\eta) \rvert $$
where the parameters have their measured values reported by particle data group \cite{pdg14} as
\begin{equation}
\lambda = 0.22537\pm 0.00061, \;\;\;\; A = 0.814^{+0.023}_{-0.024}
\nonumber
\end{equation}
\begin{equation}
\rho = 0.117\pm 0.021, \;\;\;\; \eta = 0.353\pm 0.013
\end{equation}
Thus the quark mixing angles are very small and show a hierarchy $s^{\text{CKM}}_{12} = \lambda, s^{\text{CKM}}_{23} \approx \lambda^2, s^{\text{CKM}}_{13} \approx \lambda^4$. This hierarchy among the mixing angles comes very close to the hierarchy of the quark masses and could hint towards an underlying relation between them. Several earlier works \cite{quark1, quark10, quark11, quark12, quark13, quark14} have been dedicated to relating quark mixing angles to the quark mass hierarchies.

On the other hand, neutrino masses which remain vanishing in the SM, have been found to be tiny but non-zero by several neutrino experiments \cite{PDG, PDG0, PDG1, PDG2, T2K, chooz, daya, reno}. The neutrino masses are found to obey a weaker hierarchy compared to the charged fermion masses. This can be seen from the global fit data of neutrino mass squared differences that have appeared in the recent analysis of \cite{schwetz14} and \cite{valle14}, listed in table \ref{tab:data1} and \ref{tab:data2} respectively.
\begin{center}
\begin{table}[htb]
\begin{tabular}{|c|c|c|}
\hline
Parameters & Normal Hierarchy (NH) & Inverted Hierarchy (IH) \\
\hline
$ \frac{\Delta m_{21}^2}{10^{-5} \text{eV}^2}$ & $7.02-8.09$ & $7.02-8.09 $ \\
$ \frac{|\Delta m_{3l}^2|}{10^{-3} \text{eV}^2}$ & $2.317-2.607$ & $2.307-2.590$ \\
$ \sin^2\theta_{12} $ &  $0.270-0.344$ & $0.270-0.344 $ \\
$ \sin^2\theta_{23} $ & $0.382-0.643$ &  $0.389-0.644 $ \\
$\sin^2\theta_{13} $ & $0.0186-0.0250$ & $0.0188-0.0251 $ \\
$ \delta_{CP} $ & $0-2\pi$ & $0-2\pi$ \\
\hline
\end{tabular}
\caption{Global fit $3\sigma$ values of neutrino oscillation parameters \cite{schwetz14}. Here $\Delta m_{3l}^2=\Delta m_{31}^2 >0$ for NH and $\Delta m_{3l}^2=\Delta m_{32}^2<0$ for IH.}
\label{tab:data1}
\end{table}
\end{center}
\begin{center}
\begin{table}[htb]
\begin{tabular}{|c|c|c|}
\hline
Parameters & Normal Hierarchy (NH) & Inverted Hierarchy (IH) \\
\hline
$ \frac{\Delta m_{21}^2}{10^{-5} \text{eV}^2}$ & $7.11-8.18$ & $7.11-8.18$ \\
$ \frac{|\Delta m_{31}^2|}{10^{-3} \text{eV}^2}$ & $2.30-2.65$ & $2.20-2.54 $ \\
$ \sin^2\theta_{12} $ &  $0.278-0.375$ & $0.278-0.375 $ \\
$ \sin^2\theta_{23} $ & $0.392-0.643$ &  $0.403-0.640 $ \\
$\sin^2\theta_{13} $ & $0.0177-0.0294$ & $0.0183-0.0297 $ \\
$ \delta_{CP} $ & $0-2\pi$ & $0-2\pi$ \\
\hline
\end{tabular}
\caption{Global fit $3\sigma$ values of neutrino oscillation parameters \cite{valle14}}
\label{tab:data2}
\end{table}
\end{center}
The above tables not only show weaker mass hierarchy among neutrinos, but also show how different leptonic mixing angles are from quark mixing angles. The leptonic mixing angles, at least two of them: the solar and atmospheric mixing angles $\theta_{12}, \theta_{23}$ are found to be unexpectedly large and do not seem to have any straightforward connection to the charged lepton mass hierarchies. 

It may be possible that quark and lepton mixing have completely different dynamical origin. A slightly old review of quark and lepton mixing can be found in \cite{fritxingrev}. It is however more predictive and economical if a common framework can explain these two different mixing together. Motivated by the presence of a grand unified theory at a very high energy scale, several recent as well as past works have discussed a common framework to understand the origin of quark and lepton mixing together. Some of these works can be found in \cite{QLrecent, QLrecent0, QLrecent1, QLrecent2, QLrecent3, fritxing, fritxing0, fritxing1, fritxing2, fritxing3, fritxing4, hollik1,nimai1} and references therein. In this letter, we adopt the framework used earlier in the works \cite{fritxing, fritxing0, fritxing1, fritxing2, fritxing3, fritxing4, hollik1, nimai1} to relate the lepton mixing angles with lepton mass hierarchy. Since quark mixing angles are closely related to the quark mass hierarchy, similar relation in the leptonic sector will put quarks and leptons on equal footing, which is expected in a unified theory of all fermions. Similar to these earlier works, here also we adopt a model independent approach and start with the conjecture that mixing angles in lepton sectors are related to the lepton mass ratios in the following way
\begin{equation}
\sin{\theta_{ij}} = \sqrt{\frac{m_i}{m_j}}
\label{eq:eq1}
\end{equation}
where $\theta_{ij}$ is the mixing angle either in charged lepton or neutrino sector and $m_{i,j}$ are masses of leptons such that $m_i < m_j$.

We show that, if all three mixing angles in charged lepton sector are related to the charged lepton mass ratios and all three mixing angles in the neutrino sector are related to the neutrino mass ratios as in equation \eqref{eq:eq1}, then correct values of all three leptonic mixing angles \footnote{Although both charged leptons and neutrinos belong to the lepton category, by leptonic mixing angles we mean the angles observed in neutrino oscillations. Whereas, mixing angles restricted to only charged lepton and only neutrino sectors are associated with their respective mass matrices and are not observed separately in oscillation experiments.} can not be generated simultaneously for all ranges of lightest neutrino mass. We show that correct leptonic mixing angles can be generated within their $3\sigma$ allowed ranges, if two of the mixing angles in each of charged lepton and neutrino sectors are related to the respective mass ratio in a way shown in equation \eqref{eq:eq1}, leaving the third mixing angles in both the sectors as free parameters. This also constrains the charged lepton as well as neutrino mass matrices to some specific structures, which can be written in terms of the third mixing angle and the lightest neutrino mass. We then constrain the free parameters namely, the lightest neutrino mass and the difference between the third mixing angles in the two sectors from the requirement of producing correct leptonic mixing angles. We also show how to realize some specific texture zeros in the neutrino mass matrix within this framework.

This letter is organized as follows. In section \ref{sec:sec2}, we outline and describe the formalism we adopt. In section \ref{conclude} we briefly mention our results and conclude. 

\section{Formalism}
\label{sec:sec2}
The $3\times 3$ unitary mixing matrix associated with either quark or lepton mixing can be parameterized as a product of three rotation matrices in three different planes
\begin{equation}
U = R_{23}R_{13}R_{12} 
\label{eq:U}
\end{equation}
where 
$$ R_{23}=\left( \begin{array}{ccc}
              1 & 0 & 0   \\
              0 & c_{23} & s_{23} \\
              0 & -s_{23} & c_{23}
                      \end{array} \right), \;\; R_{13}=\left( \begin{array}{ccc}
              c_{13} & 0 & s_{13}e^{-i\delta}   \\
              0 & 1 & 0 \\
              -s_{13}e^{i\delta} & 0 & c_{13}
                      \end{array} \right), \;\; R_{12}=\left( \begin{array}{ccc}
              c_{12} & s_{12} & 0   \\
              -s_{12} & c_{12} & 0 \\
              0 & 0 & 1
                      \end{array} \right)$$

where $c_{ij} = \cos{\theta_{ij}}, s_{ij} = \sin{\theta_{ij}}$ and $\delta$ is the Dirac CP phase. The CKM mixing matrix in the quark sector can be written as
\begin{equation}
V_{\text{CKM}} = V^{u\dagger}_L V^{d}_L
\end{equation}
where $V^u_L, V^d_L$ are two of the four unitary matrices required to diagonalize the up and down type quark mass matrices. Expressing both of these mixing matrices as products of three rotation matrices shown in equation \eqref{eq:U}, one can express the CKM matrix as
\begin{equation}
V_{\text{CKM}} = R^{u\dagger}_{12}R^{u\dagger}_{13}R^{u\dagger}_{23}R^d_{23}R^d_{13}R^d_{12}
\end{equation}
Since quark mixing angles are closely related to quark mass hierarchies, assuming negligible $1-3$ mixing (due to strong hierarchy between first and third generation quark masses), a convenient representation of CKM matrix was suggested by \cite{fritxingrev,fritxing20, fritxing21}
\begin{equation}
V_{\text{CKM}} = R^{u\dagger}_{12}R_{23}R^d_{12}
\end{equation}
where $R_{23} = R^{u\dagger}_{23}R^d_{23}$ can be parameterized in terms of an angle $\theta$ and a phase $\phi$. The rotation matrices $R^{u}_{12}, R^d_{12}$ are parameterized by angle
$$ \sin{\theta^u_{12}} \approx \tan{\theta^u_{12}} = \sqrt{\frac{m_u}{m_c}} $$
$$ \sin{\theta^d_{12}} \approx \tan{\theta^d_{12}} = \sqrt{\frac{m_d}{m_s}} $$
\footnote{Due to strong mass hierarchies in the charged fermion sector one can use $\sin{\theta}$ and $\tan{\theta}$ interchangeably. Difference may arise in the neutrino sector due to weaker mass hierarchies, we do not get correct result using $\tan{\theta}$ in the leptonic sector and hence adopt $\sin{\theta}$ in rest of our work.}
which is very close to the observed pattern of quark masses and mixing and encourages one to look into the lepton sector carefully and check whether such a relation between mixing angles and mass hierarchy exists.

The Pontecorvo-Maki-Nakagawa-Sakata (PMNS) leptonic mixing matrix is related to the diagonalizing 
matrices of neutrino and charged lepton mass matrices $U_{\nu}, U_{\ell}$ respectively, as
\begin{equation}
U_{\text{PMNS}} = U^{\dagger}_{\ell} U_{\nu}
\label{pmns0}
\end{equation}
Using a similar approach discussed above in the case of quarks, the authors of \cite{fritxing} proposed a parameterization of leptonic mixing given by 
\begin{equation}
U_{\text{PMNS}} = \left( \begin{array}{ccc}
              c_l & s_l & 0   \\
              -s_l & c_l & 0 \\
              0 & 0 & 1
                      \end{array} \right)\left( \begin{array}{ccc}
              1 & 0 & 0   \\
              0 & c & s \\
              0 & -s & c
                      \end{array} \right)\left(\begin{array}{ccc}
              c_{\nu} & -s_{\nu} & 0   \\
              s_{\nu} & c_{\nu} & 0 \\
              0 & 0 & 1
                      \end{array} \right)
\label{Upmnspara1}
\end{equation}
with $c_l = \cos{\theta_l}, s_l = \sin{\theta_l}, c_{\nu} = \cos{\theta_{\nu}}, s_{\nu} = \sin{\theta_{\nu}}, c = \cos{\theta}, s = \sin{\theta}$. Here we are ignoring the CP phases, both Dirac and Majorana. Assuming solar and atmospheric neutrino oscillations to be decoupled, the three leptonic mixing angles observed in neutrino oscillations are given by
$$ \theta_{12} \approx \theta_{\nu}, \;\; \theta_{23} \approx \theta, \;\; \theta_{13} \approx \theta_l \sin{\theta} $$
The authors further conjectured that the mixing angles $\theta_l, \theta_{\nu}$ could be related to the lepton mass ratios $m_e/m_{\mu}, m_1/m_2$ in a way given in equation \eqref{eq:eq1}. The smallness of $\theta_{13}$ therefore, was attributed to the strong hierarchy in the charged lepton sector. However, the value of $\theta_{13}$ in this framework came out to be very small, around $3^o$, which is ruled out by present neutrino data. We therefore, do not assume the parameterization of $U_{\text{PMNS}}$ to be similar to $V_{\text{CKM}}$, given by equation \eqref{Upmnspara1} in this work. This is mainly due to the fact that hierarchy in the lepton sector (neutrinos, in particular) are very different from the quark sector and hence the assumptions used in the derivation of the above parameterization of $V_{\text{CKM}}$ need not be valid for $U_{\text{PMNS}}$.

Instead, we consider a general analysis first where all mixing angles in charged lepton as well as neutrino sector are related to the corresponding mass hierarchies in a way shown by equation \eqref{eq:eq1}. This help us to determine the structure of charged lepton mass matrix (assuming to be symmetric) completely and the neutrino mass matrix in terms of one free parameter, the lightest neutrino mass. These mass matrices, in general contain CP phases which we are ignoring in our work as they are not going to affect our main results substantially. After showing that this formalism does not give rise to the correct leptonic mixing angles, we then assume the validity of the relation between mixing angles and mass hierarchies \eqref{eq:eq1} to be restricted to only two mixing angles in each of charged lepton and neutrino sector, whereas the third mixing angle in both the sectors are free parameters. We show that this scenario can give rise to the leptonic mixing angles within their $3\sigma$ range. This formalism allows us to construct the symmetric charged lepton mass matrix upto one free parameter, and the neutrino mass matrix upto two free parameters.
\begin{figure}[ht]
\begin{center}
$
\begin{array}{cc}
\includegraphics[width=0.5\textwidth]{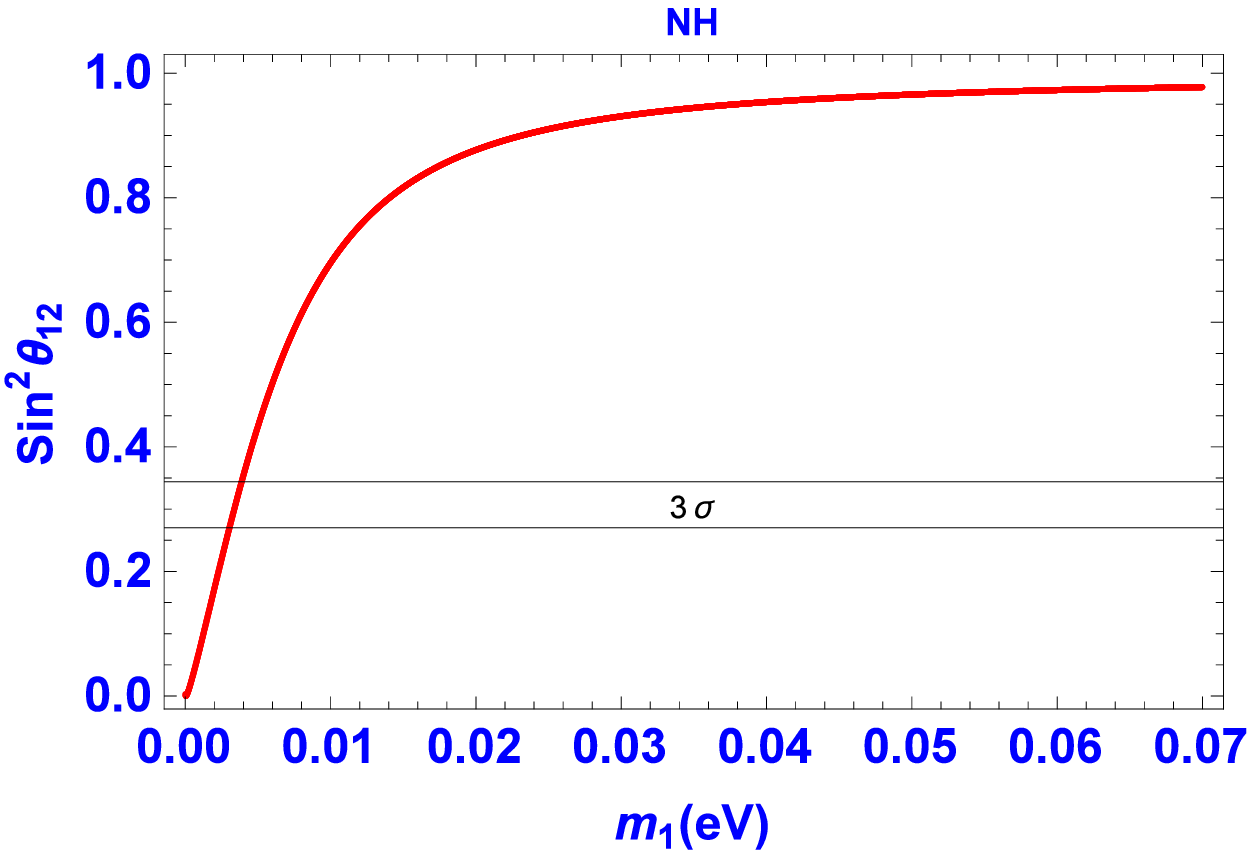} &
\includegraphics[width=0.5\textwidth]{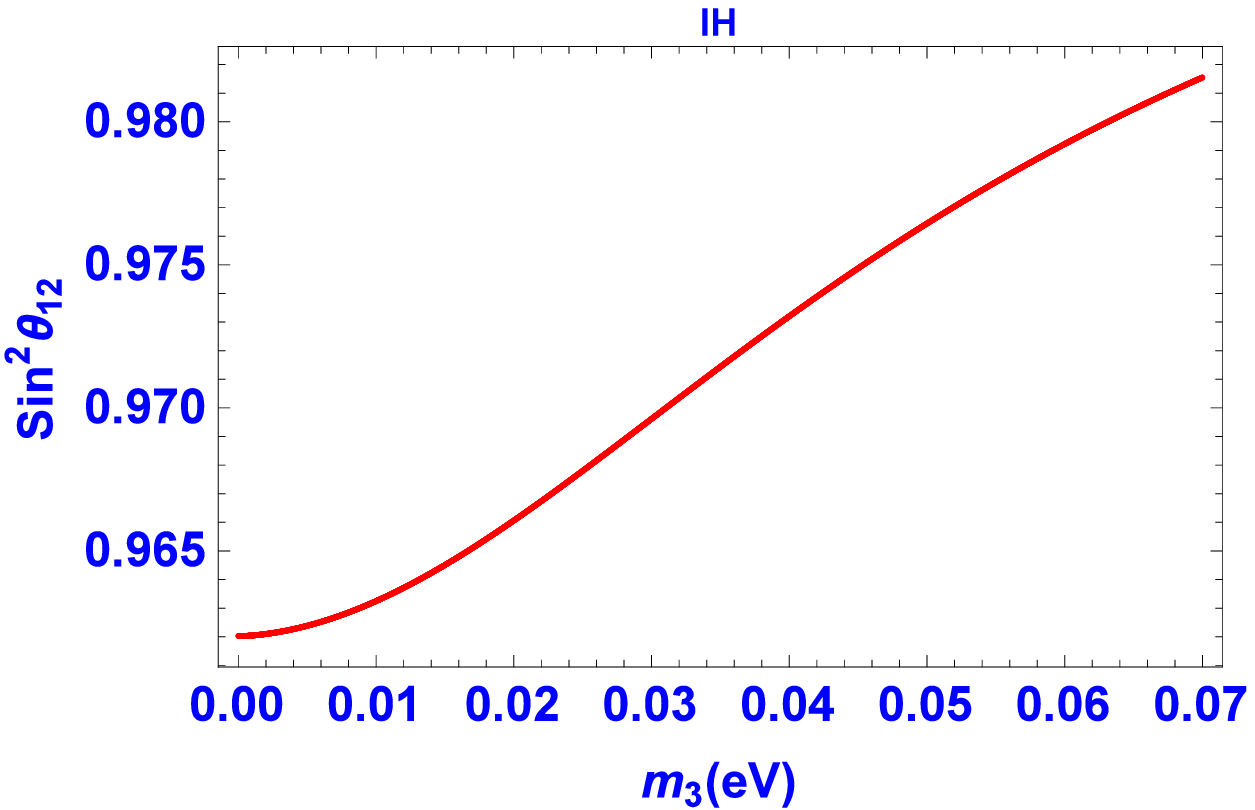} \\
\includegraphics[width=0.5\textwidth]{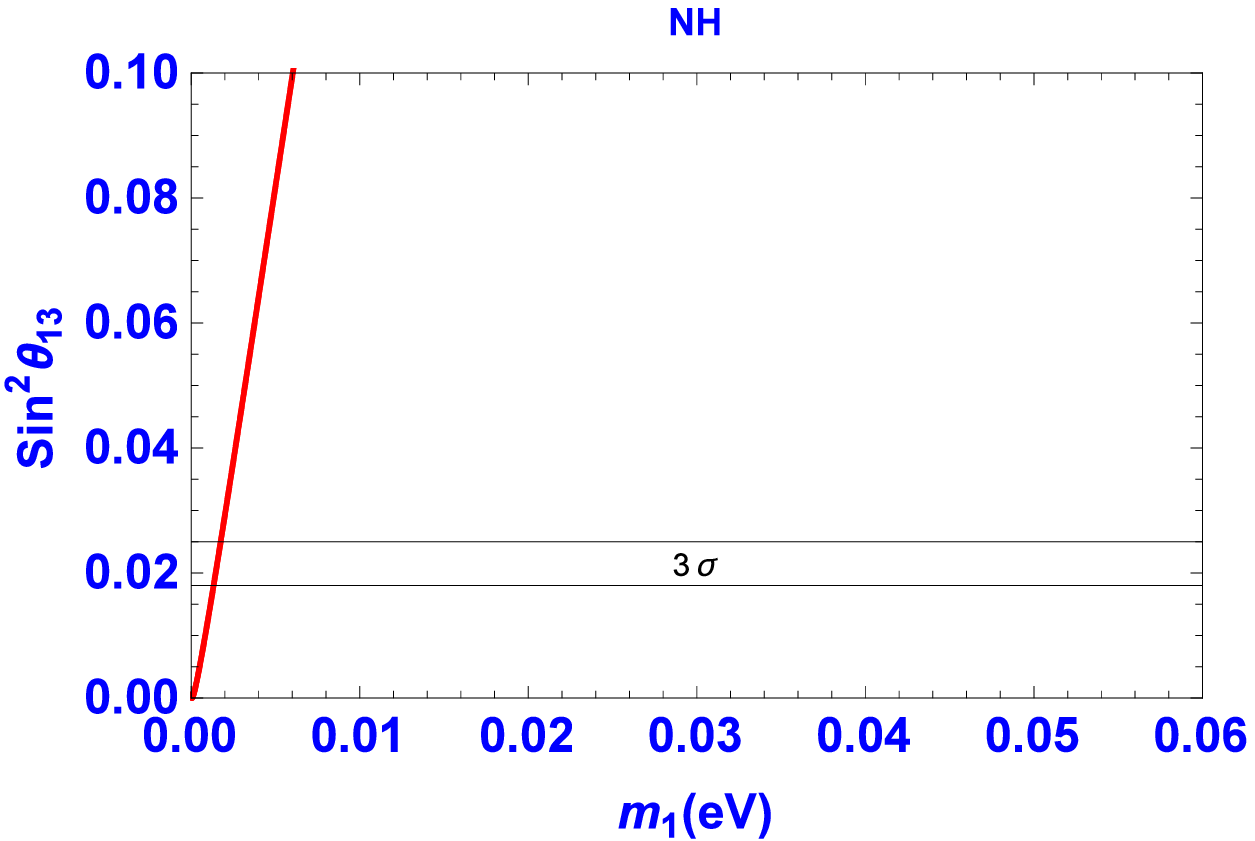} &
\includegraphics[width=0.5\textwidth]{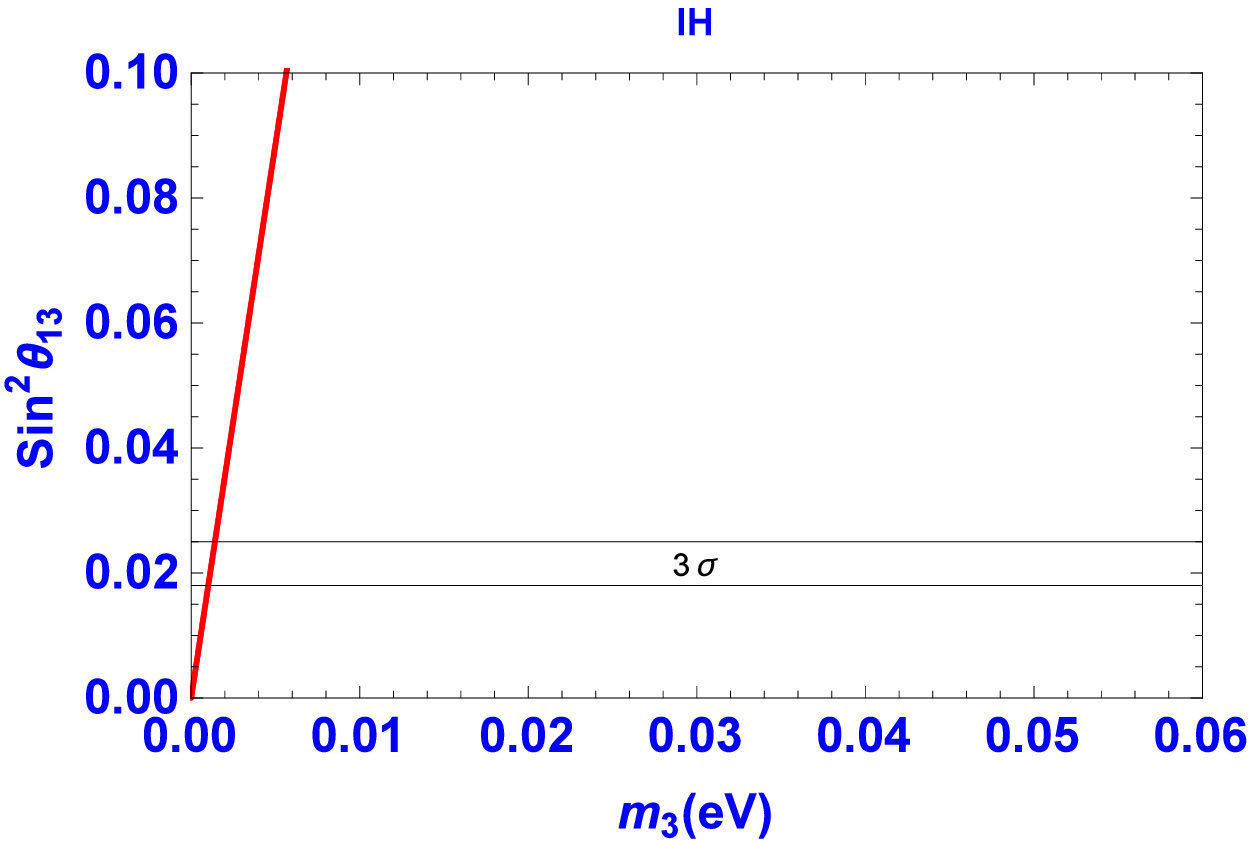} \\
\includegraphics[width=0.5\textwidth]{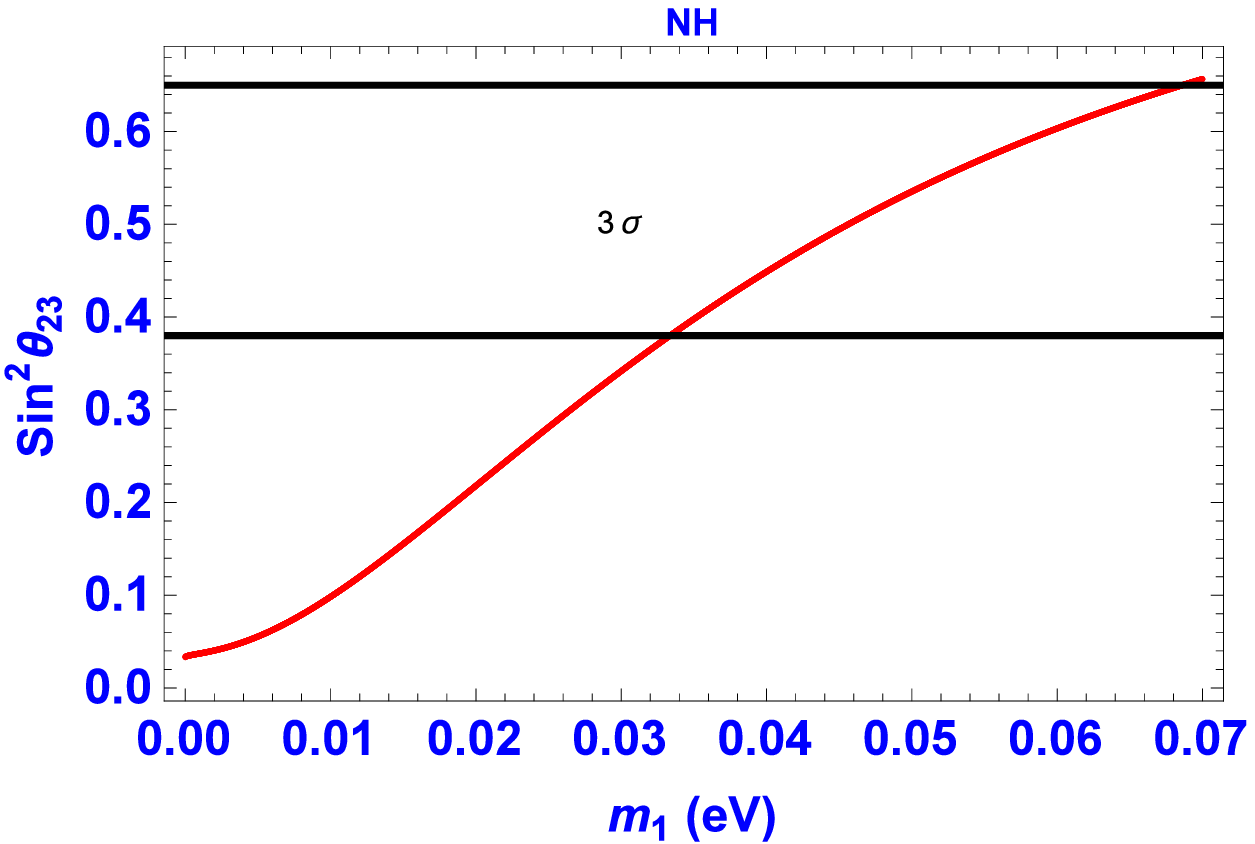} &
\includegraphics[width=0.5\textwidth]{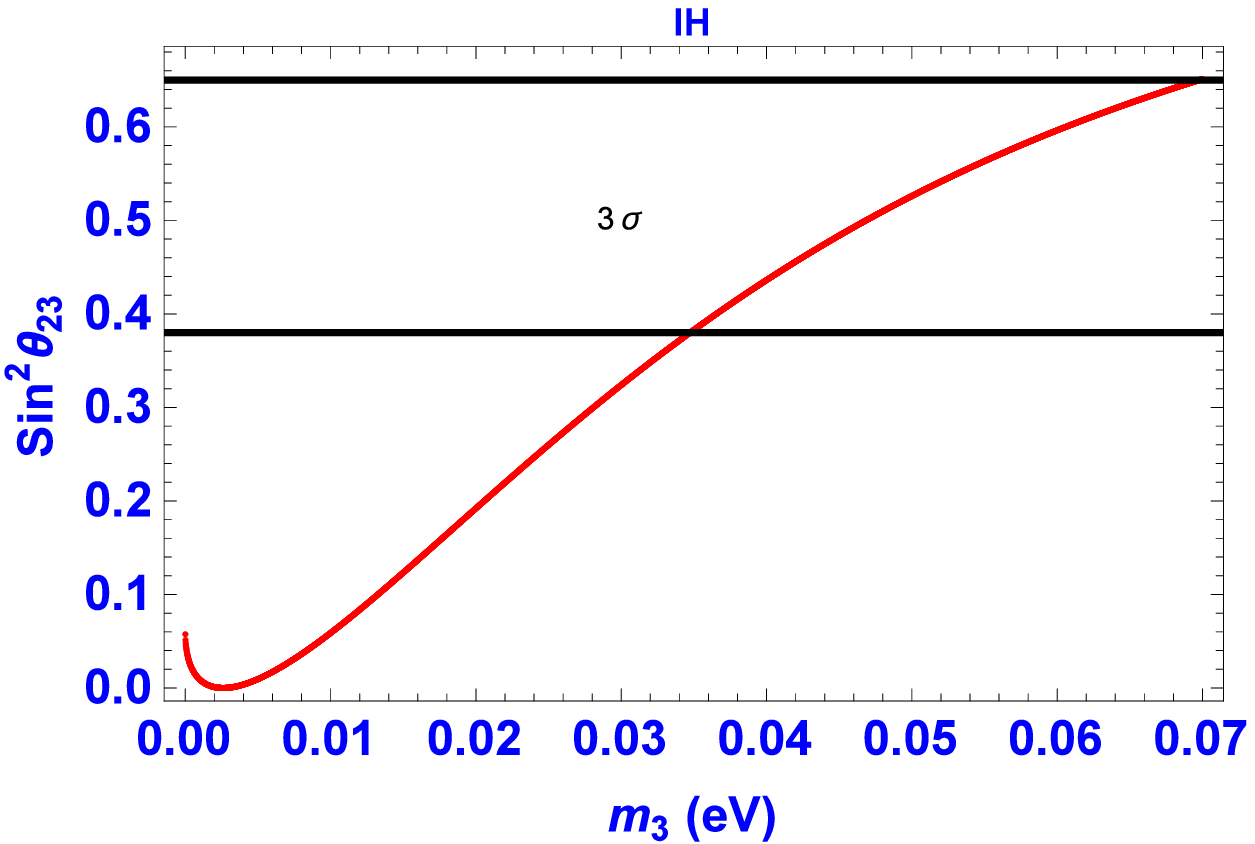}
\end{array}$
\end{center}
\caption{Leptonic mixing angles as functions of lightest neutrino mass for both normal and inverted hierarchical neutrino masses in scenario I}
\label{fig1}
\end{figure}
\subsection{Scenario I}
The PMNS mixing matrix shown in equation \eqref{pmns0} can be written as
\begin{equation}
U_{\text{PMNS}} = R^{l\dagger}_{12}R^{l\dagger}_{13}R^{l\dagger}_{23}R^{\nu}_{23}R^{\nu}_{13}R^{\nu}_{12}
\label{pmns1}
\end{equation}
The mixing angles in the charged lepton sector rotation matrices are given by 
$$ \sin{\theta^l_{12}} = \sqrt{\frac{m_e}{m_{\mu}}}, \;\;  \sin{\theta^l_{13}} = \sqrt{\frac{m_e}{m_{\tau}}}, \;\;  \sin{\theta^l_{23}} = \sqrt{\frac{m_{\mu}}{m_{\tau}}} $$
In the neutrino sector with normal hierarchy, the mixing angles are 
$$ \sin{\theta^{\nu}_{12}} = \sqrt{\frac{m_1}{m_{2}}}, \;\;  \sin{\theta^{\nu}_{13}} = \sqrt{\frac{m_1}{m_{3}}}, \;\;  \sin{\theta^{\nu}_{23}} = \sqrt{\frac{m_2}{m_3}} $$
On the other hand, for inverted hierarchical neutrino mass, we assume the mixing angles to be 
$$ \sin{\theta^{\nu}_{12}} = \sqrt{\frac{m_1}{m_{2}}}, \;\;  \sin{\theta^{\nu}_{13}} = \sqrt{\frac{m_3}{m_{1}}}, \;\;  \sin{\theta^{\nu}_{23}} = \sqrt{\frac{m_3}{m_2}} $$
Thus, the PMNS mixing matrix given in equation \eqref{pmns1} can be constructed in terms of the lepton mass ratios. For the case of normal hierarchy, the three neutrino masses can be written as $(m_1, \sqrt{m^2_1+\Delta m_{21}^2}, \sqrt{m_1^2+\Delta m_{31}^2})$, while for the case of inverted hierarchy, they can be written as 
$(\sqrt{m_3^2+\Delta m_{23}^2-\Delta m_{21}^2}, \sqrt{m_3^2+\Delta m_{23}^2}, m_3)$. We use the charged lepton masses from particle data group data given in \cite{pdg14}, best fit values of neutrino mass squared differences from \cite{schwetz14} leaving the lightest neutrino mass as free parameter. The leptonic mixing angles can be calculated from the PMNS mixing matrix as
$$ \theta_{13} = \text{arcsin} \left ( \lvert (U_{\text{PMNS}})_{13} \rvert \right ) $$
$$ \theta_{12} = \text{arctan} \left ( \frac{\lvert (U_{\text{PMNS}})_{12} \rvert }{\lvert (U_{\text{PMNS}})_{11} \rvert} \right) $$
\begin{equation}
\theta_{23} = \text{arctan} \left ( \frac{\lvert (U_{\text{PMNS}})_{23} \rvert }{\lvert (U_{\text{PMNS}})_{33} \rvert} \right)
\label{mixang}
\end{equation}
We then calculate the the leptonic mixing angles as a function of lightest neutrino mass. As can be seen from the plots, all the leptonic mixing angles can not be generated simultaneously for both normal and inverted hierarchies within this framework. Therefore, we do not pursue this any further. It should be noted that we have not taken the leptonic Dirac CP phase $\delta_{CP}$ into account in this analysis. The general conclusion we arrive in this case however, does not depend on a particular value of $\delta_{CP}$.

\begin{figure}[ht]
\begin{center}
$
\begin{array}{cc}
\includegraphics[width=0.5\textwidth]{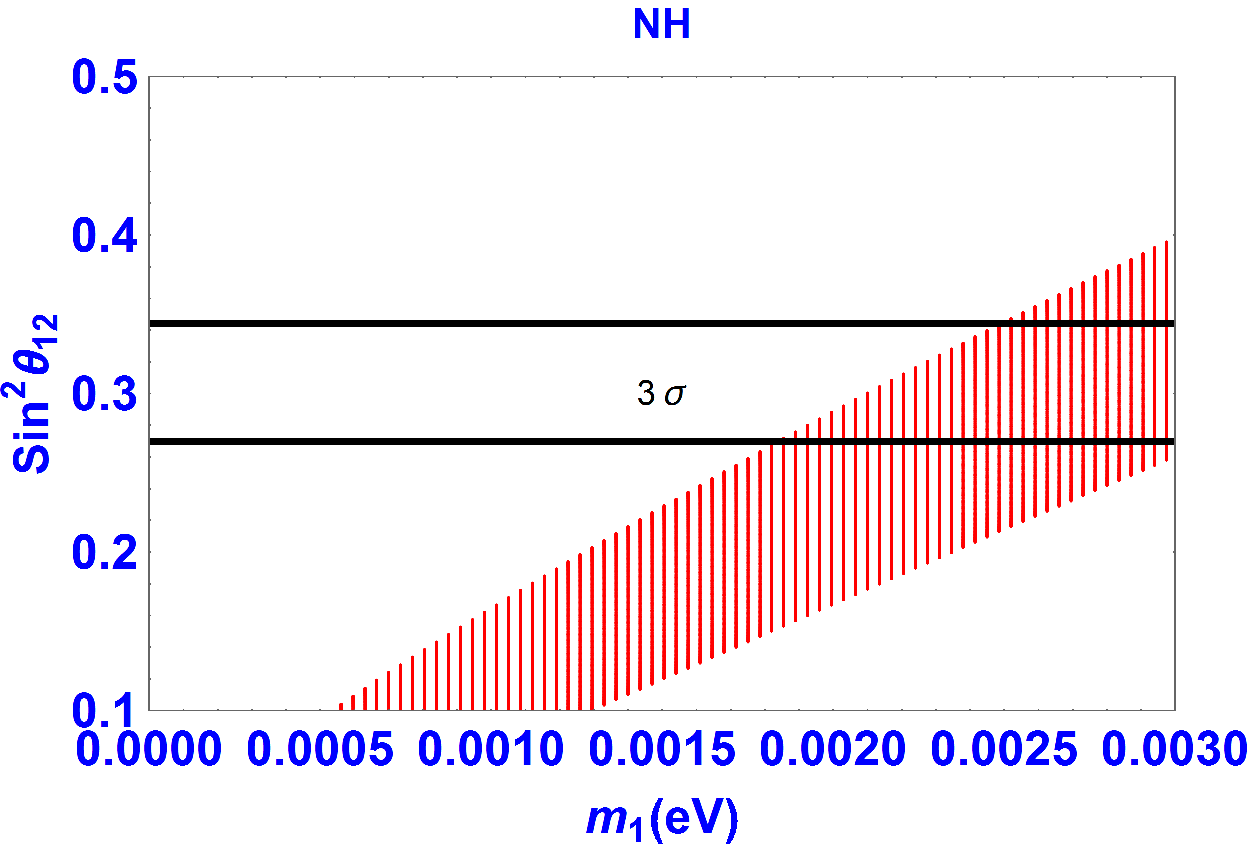} &
\includegraphics[width=0.5\textwidth]{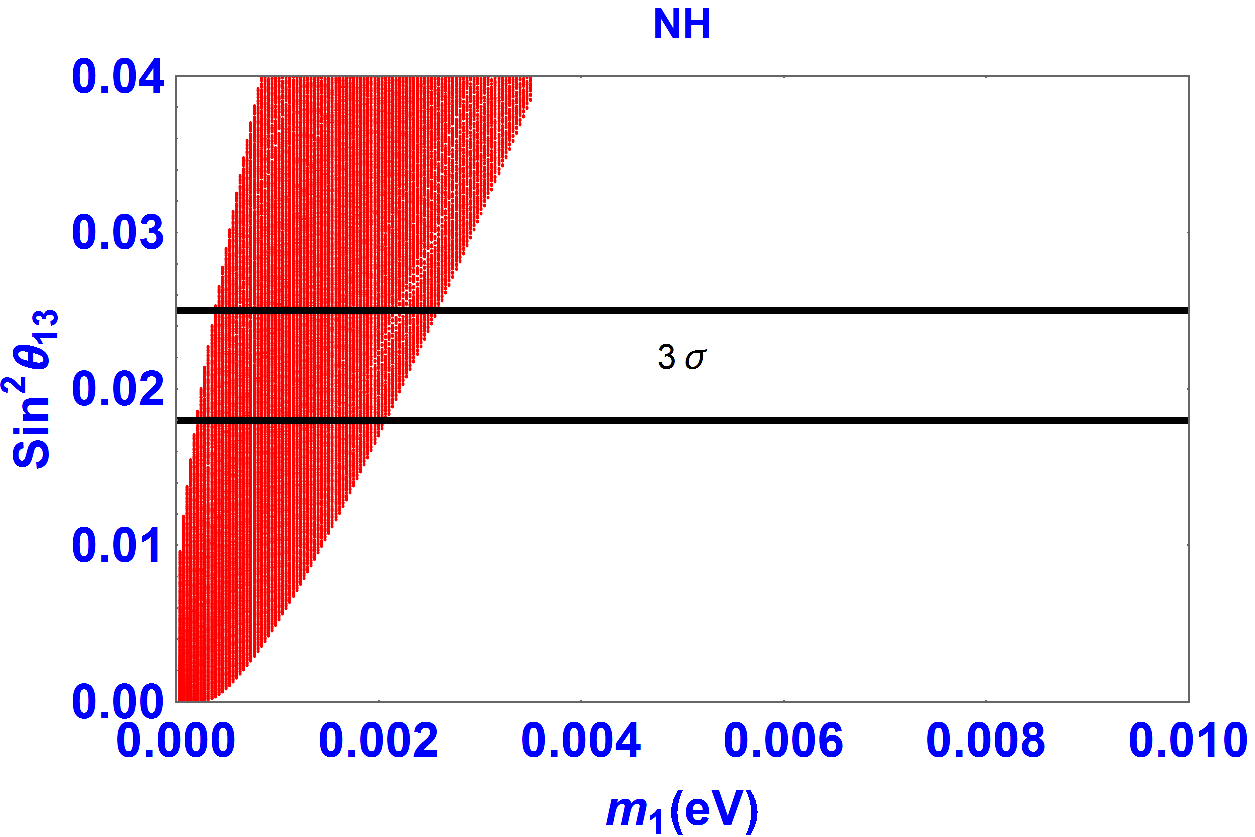} \\
\includegraphics[width=0.5\textwidth]{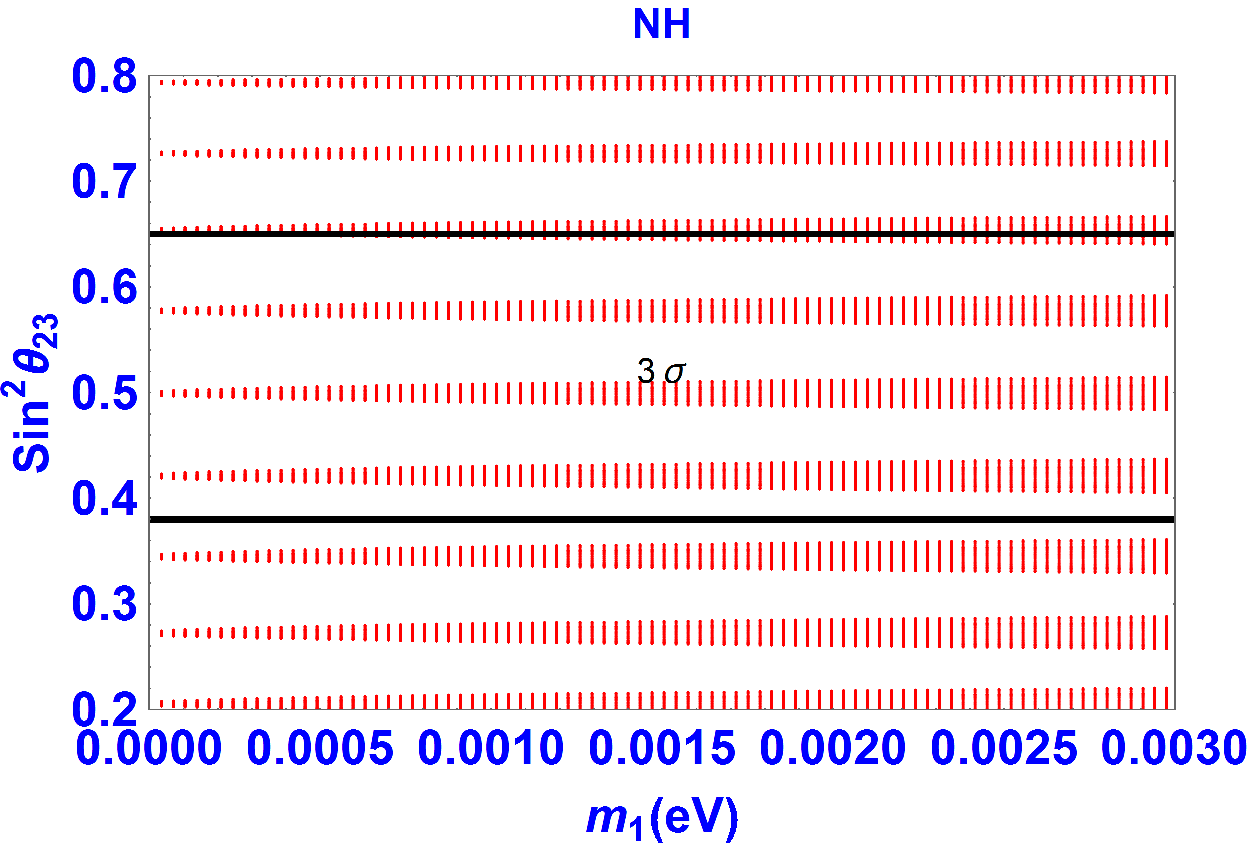} &
\includegraphics[width=0.5\textwidth]{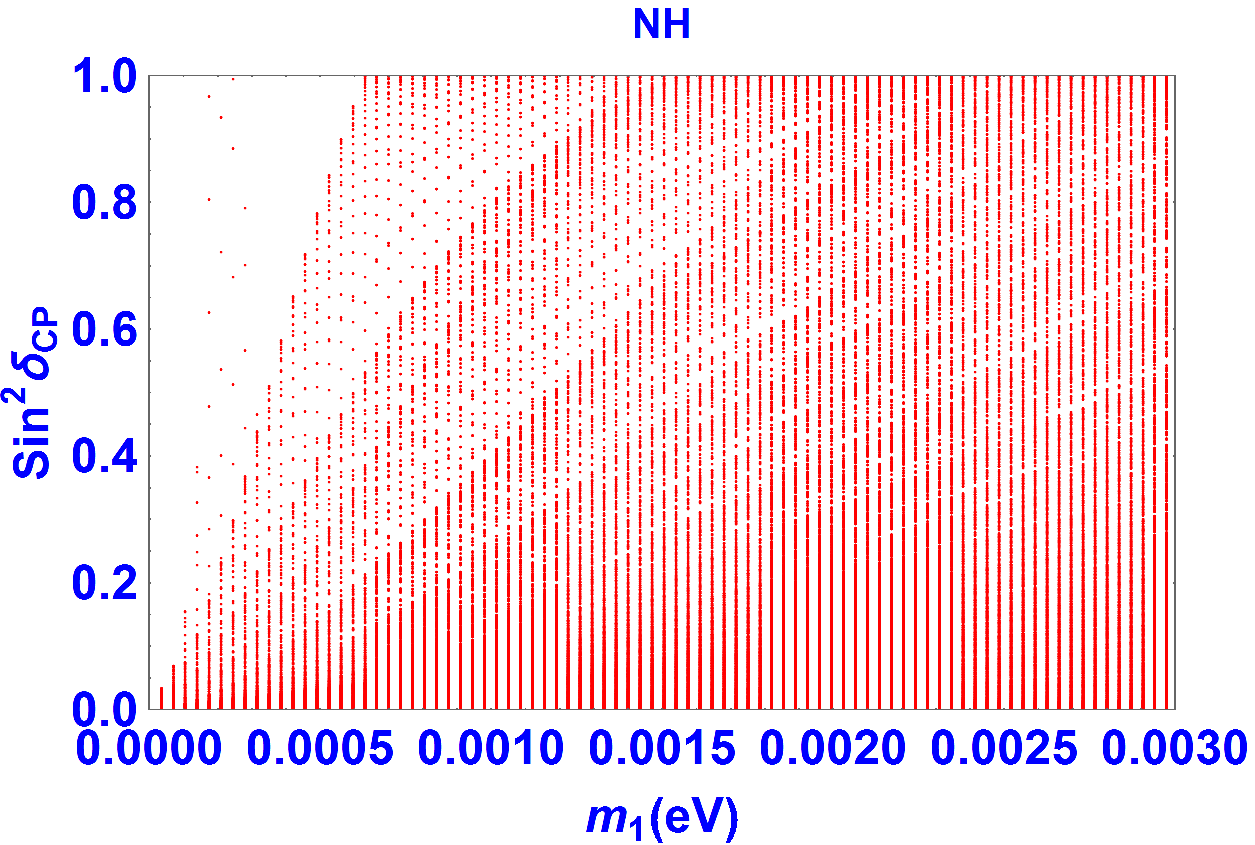}
\end{array}$
\end{center}
\caption{Leptonic mixing angles $\sin^2{\theta_{12}}, \sin^2{\theta_{13}}, \sin^2{\theta_{23}}$ and Dirac CP phase $\sin^2{\delta_{CP}}$ in terms of lightest neutrino mass $m_1$ in scenario II}
\label{fig2}
\end{figure}
\begin{figure}[ht]
\begin{center}
$
\begin{array}{cc}
\includegraphics[width=0.5\textwidth]{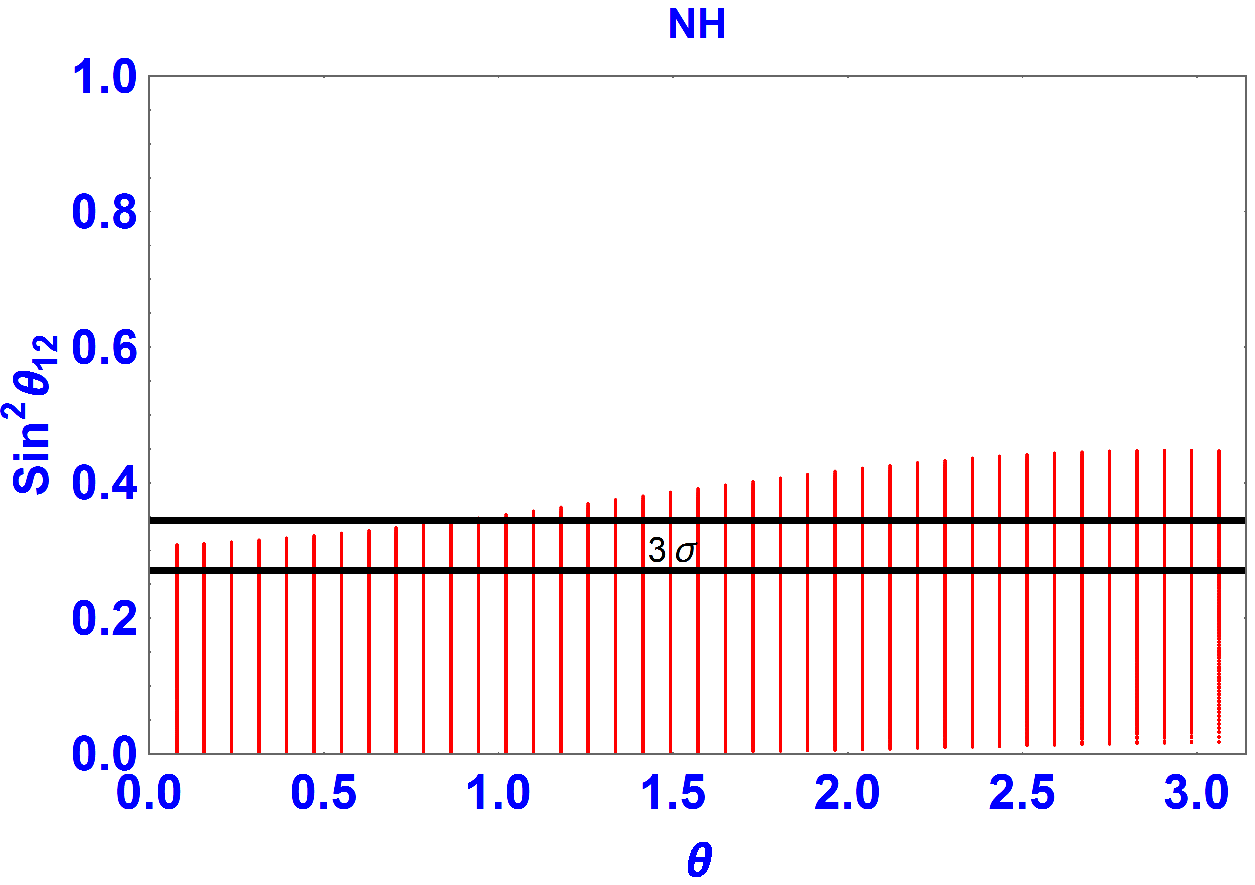} &
\includegraphics[width=0.5\textwidth]{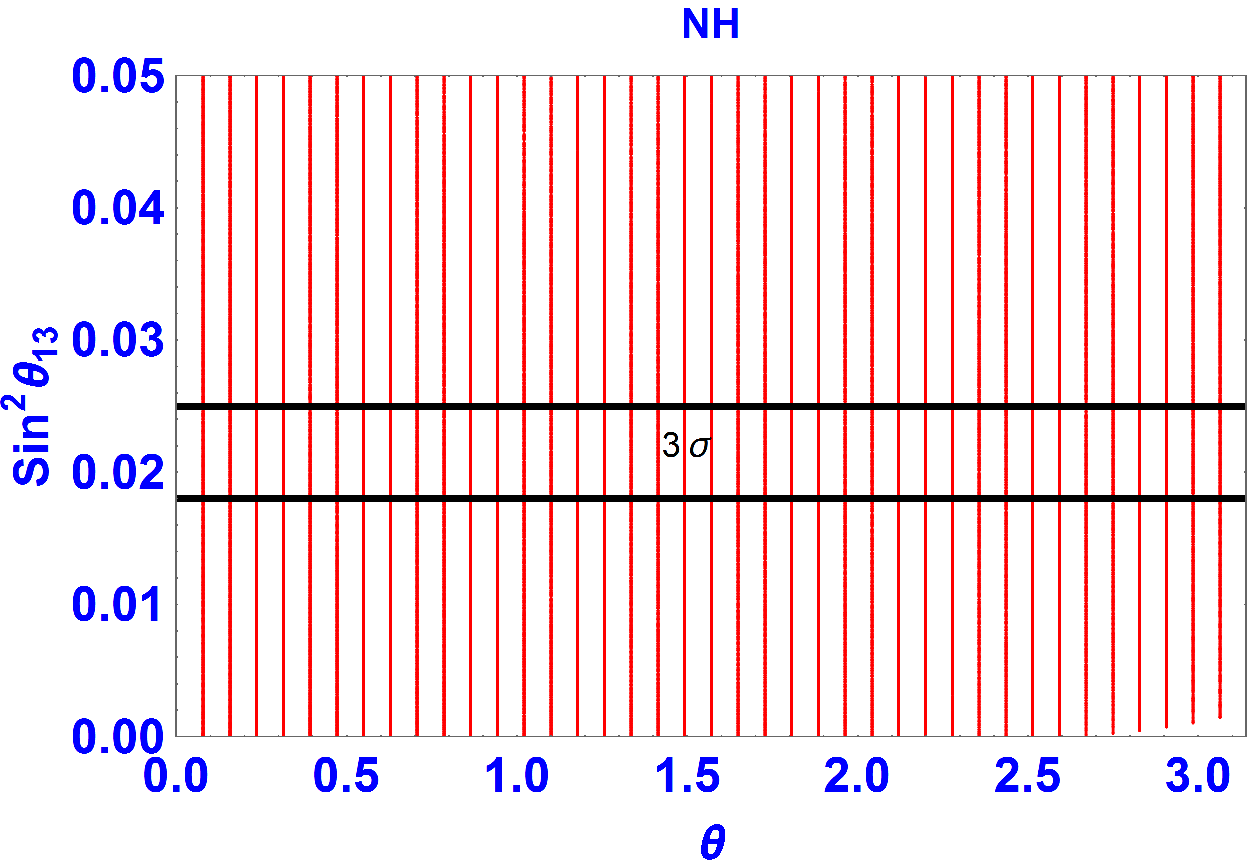} \\
\includegraphics[width=0.5\textwidth]{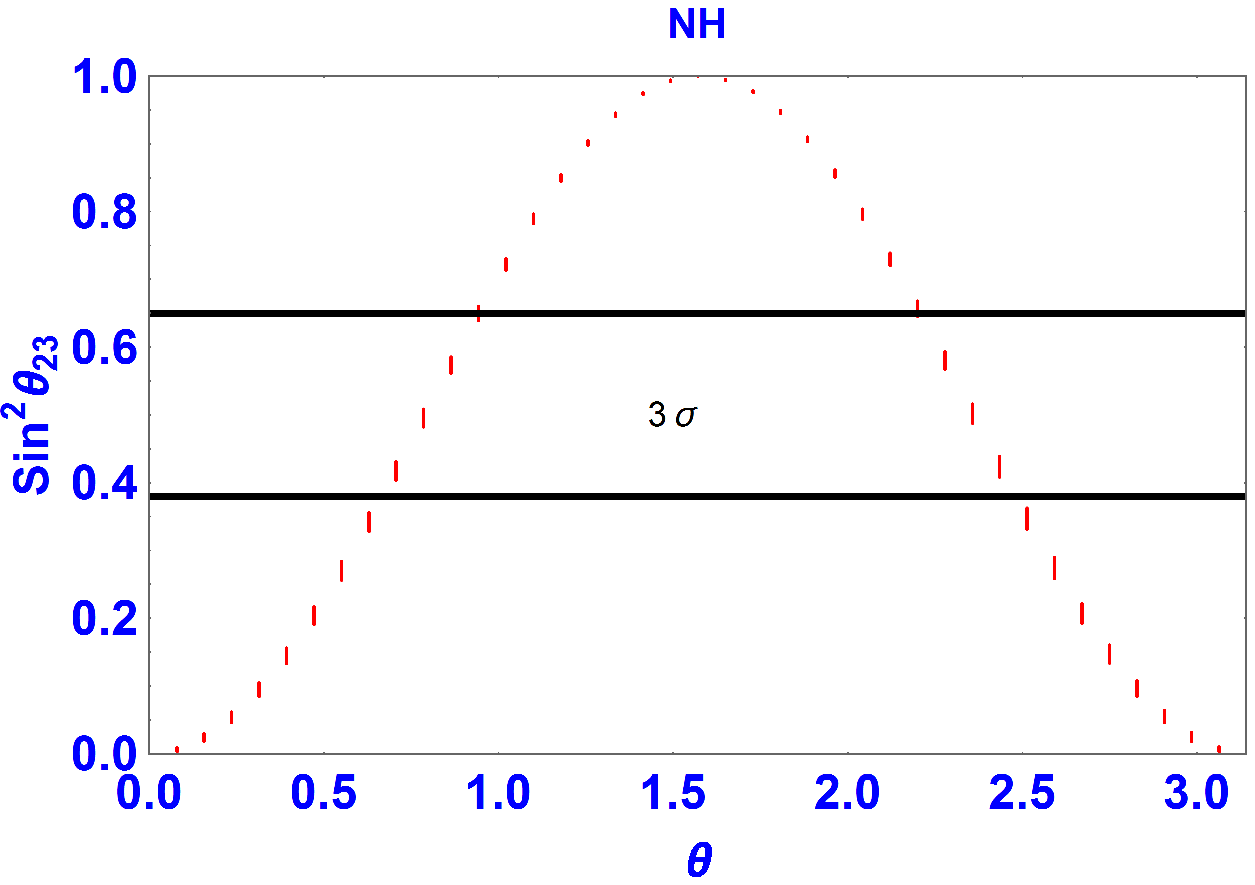} &
\includegraphics[width=0.5\textwidth]{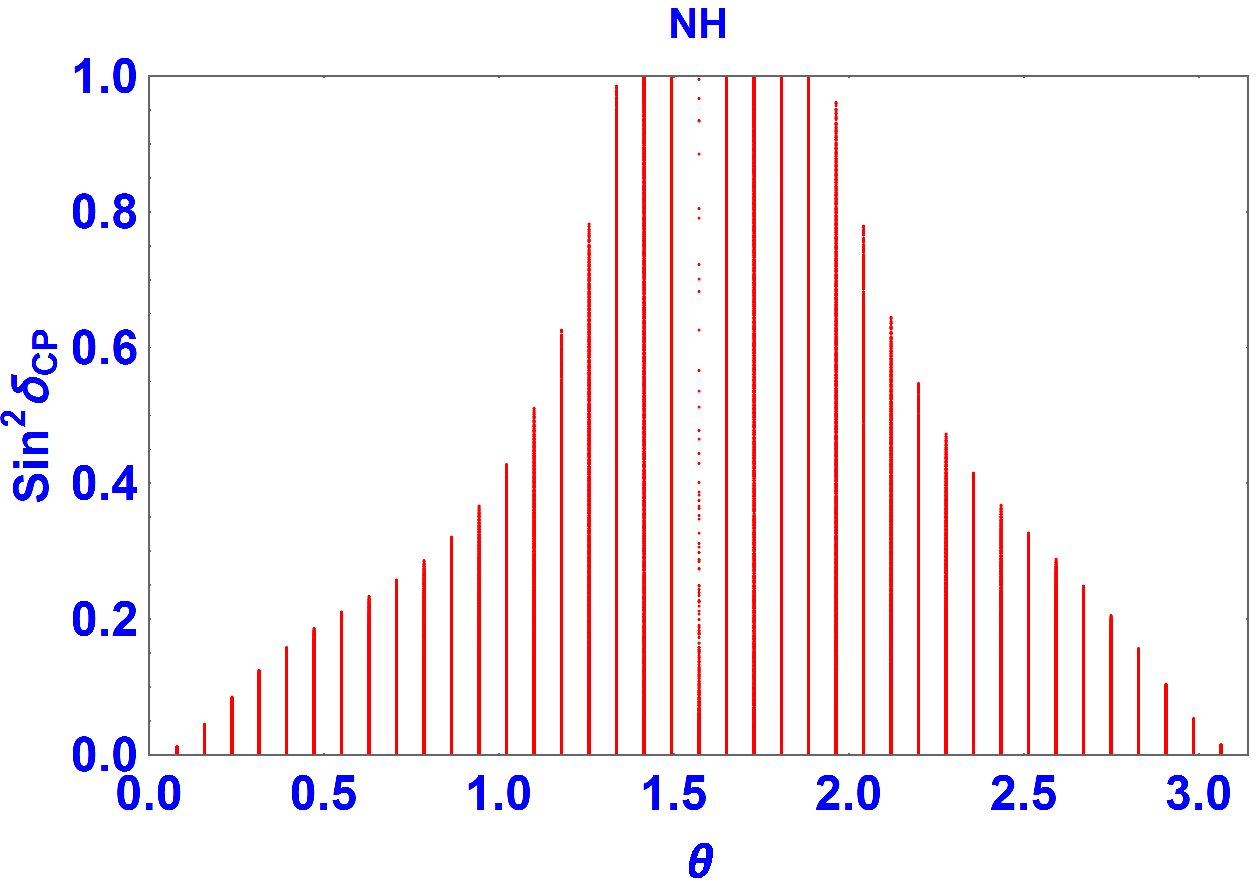}
\end{array}$
\end{center}
\caption{Leptonic mixing angles $\sin^2{\theta_{12}}, \sin^2{\theta_{13}}, \sin^2{\theta_{23}}$ and Dirac CP phase $\sin^2{\delta_{CP}}$ in terms of $\theta = \theta^{\nu}_{23}-\theta^l_{23}$ in scenario II}
\label{fig3}
\end{figure}

\subsection{Scenario II}
Instead of assuming all the mixing angles in both charged lepton and neutrino sector to be related to the lepton mass hierarchies as discussed above, we now assume such a relation to be valid only for two mixing angles in each sector. For simplicity, we assume 1-2 and 1-3 mixing angles to be related to the corresponding mass hierarchies. Whereas the rotation matrices in the 2-3 sectors are assumed to be of the forms given below.
$$ R^l_{23}=\left( \begin{array}{ccc}
              1 & 0 & 0   \\
              0 & c^l_{23} & s^l_{23}e^{-i\delta^l} \\
              0 & -s^l_{23}e^{i\delta^l} & c^l_{23}
                      \end{array} \right), \;\;\; R^{\nu}_{23}=\left( \begin{array}{ccc}
              1 & 0 & 0   \\
              0 & c^{\nu}_{23} & s^{\nu}_{23}e^{-i\delta^{\nu}} \\
              0 & -s^{\nu}_{23}e^{\i\delta^{\nu}} & c^{\nu}_{23}
                      \end{array} \right) $$
so that $R^{l\dagger}_{23}R^{\nu}_{23}$ part of $U_{\text{PMNS}}$ can simply be written as
$$ R^{l\dagger}_{23}R^{\nu}_{23} =  \left( \begin{array}{ccc}
              1 & 0 & 0   \\
              0 & c_{\theta} & s_{\theta} e^{-i\delta} \\
              0 & -s_{\theta} e^{i\delta} & c_{\theta}
                      \end{array} \right) $$   
where $c_{\theta} = \cos{\theta}, s_{\theta} = \sin{\theta}, \theta = \theta^{\nu}_{23}-\theta^l_{23}$. Also, $\delta = \delta^l = \delta^{\nu}$ is assumed. Using these notations, the PMNS mixing matrix can be constructed with the following elements.
\small{\begin{equation}
(U_{\text{PMNS}})_{11}=\left(c_{\theta } s^l_{12}-e^{i \delta } c^l_{12} s_{\theta } s^l_{13}\right) s^{\nu }_{12}+c^{\nu }_{12} \left(e^{-i \delta } s_{\theta
} s^l_{12} s^{\nu }_{13}+c^l_{12} \left(c^l_{13} c^{\nu }_{13}+c_{\theta } s^l_{13} s^{\nu }_{13}\right)\right)
\nonumber
\end{equation}
\begin{equation}
(U_{\text{PMNS}})_{12}=c^{\nu }_{12} \left(-c_{\theta } s^l_{12}+e^{i \delta } c^l_{12} s_{\theta } s^l_{13}\right)+s^{\nu }_{12} \left(e^{-i \delta } s_{\theta
} s^l_{12} s^{\nu }_{13}+c^l_{12} \left(c^l_{13} c^{\nu }_{13}+c_{\theta } s^l_{13} s^{\nu }_{13}\right)\right)
\nonumber
\end{equation}              
\begin{equation}
(U_{\text{PMNS}})_{13}=c^{\nu }_{13} \left(-e^{-i \delta } s_{\theta } s^l_{12}-c_{\theta } c^l_{12} s^l_{13}\right)+c^l_{12} c^l_{13} s^{\nu }_{13}
\nonumber
\end{equation}
\begin{equation}
(U_{\text{PMNS}})_{21}=-\left(c_{\theta } c^l_{12}+e^{i \delta } s_{\theta } s^l_{12} s^l_{13}\right) s^{\nu }_{12}+c^{\nu }_{12} \left(c^l_{13} c^{\nu }_{13}
s^l_{12}+\left(-e^{-i \delta } c^l_{12} s_{\theta }+c_{\theta } s^l_{12} s^l_{13}\right) s^{\nu }_{13}\right)
\nonumber
\end{equation}
\begin{equation}
(U_{\text{PMNS}})_{22}=c^{\nu }_{12} \left(c_{\theta } c^l_{12}+e^{i \delta } s_{\theta } s^l_{12} s^l_{13}\right)+s^{\nu }_{12} \left(c^l_{13} c^{\nu }_{13}
s^l_{12}+\left(-e^{-i \delta } c^l_{12} s_{\theta }+c_{\theta } s^l_{12} s^l_{13}\right) s^{\nu }_{13}\right)
\nonumber
\end{equation}
\begin{equation}
(U_{\text{PMNS}})_{23}=e^{-i \delta } c^l_{12} c^{\nu }_{13} s_{\theta }+s^l_{12} \left(-c_{\theta } c^{\nu }_{13} s^l_{13}+c^l_{13} s^{\nu }_{13}\right)
\nonumber
\end{equation}
\begin{equation}
(U_{\text{PMNS}})_{31}=e^{i \delta } c^l_{13} s_{\theta } s^{\nu }_{12}+c^{\nu }_{12} \left(c^{\nu }_{13} s^l_{13}-c_{\theta } c^l_{13} s^{\nu }_{13}\right)
\nonumber
\end{equation}
\begin{equation}
(U_{\text{PMNS}})_{32}=c^{\nu }_{13} s^l_{13} s^{\nu }_{12}-c^l_{13} \left(e^{i \delta } c^{\nu }_{12} s_{\theta }+c_{\theta } s^{\nu }_{12} s^{\nu }_{13}\right)
\nonumber
\end{equation}
\begin{equation}
(U_{\text{PMNS}})_{33}=c_{\theta } c^l_{13} c^{\nu }_{13}+s^l_{13} s^{\nu }_{13}
\end{equation}}
Due to the strong charged lepton mass hierarchy, one can assume $s^l_{13} \approx \theta^l_{13}, c^l_{13} \approx 1, s^l_{12} \approx \theta^l_{12}, c^l_{12} \approx 1$. Using these approximations, one can further simplify the PMNS matrix elements. For example, the 1-3 element can now be written as
\small{\begin{widetext}
         \begin{equation}
(U_{\text{PMNS}})_{13}=\sqrt{\frac{m_1}{m_3}}+\sqrt{1-\frac{m_1}{m_3}} \left(-c_{\theta } \sqrt{\frac{m_e}{m_{\tau }}}-e^{-i \delta } \sqrt{\frac{m_e}{m_{\mu }}} s_{\theta
}\right)
\nonumber
\end{equation}
\end{widetext}}
Similar expressions can also be written for inverted hierarchy with the mass ratios appropriately related to the mixing angles. It can be seen from the above expressions that the leptonic mixing angles are functions of three free parameters namely, $\theta = \theta^{\nu}_{23}-\theta^l_{23}$, $\delta$ and the lightest neutrino mass $m_1$ (NH), $m_3$ (IH). We vary these three parameters and numerically calculate the leptonic mixing angles and leptonic Dirac CP phase. The mixing angles are calculated by using the formula given by equation \eqref{mixang} whereas the Dirac CP phase can be extracted from the PMNS matrix as \cite{Antusch}
\begin{equation}
\delta_{CP} = -\text{arg} \left ( \frac{\frac{U^*_{ii}U_{ij}U_{ji}U^*_{jj}}{c_{12}c^2_{13}c_{23}s_{13}}+c_{12}c_{23}s_{13}}{s_{12}s_{23}} \right )
\end{equation}
where $U_{ij} = (U_{\text{PMNS}})_{ij}, s_{ij} = \sin{\theta_{ij}}, c_{ij} = \cos{\theta_{ij}}$ and $i, j \varepsilon {1,2,3}, i\neq j$. 
\begin{figure}[ht]
\begin{center}
$
\begin{array}{cc}
\includegraphics[width=0.5\textwidth]{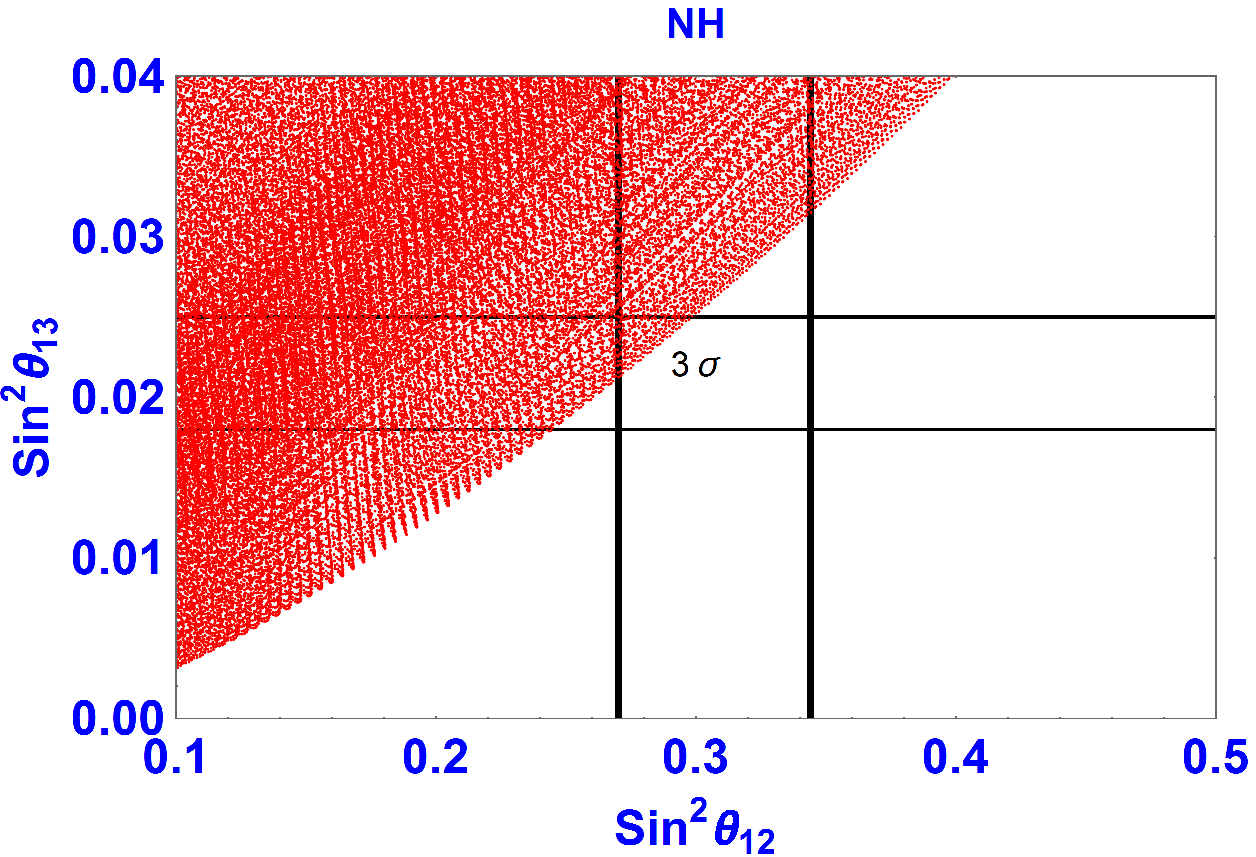} &
\includegraphics[width=0.5\textwidth]{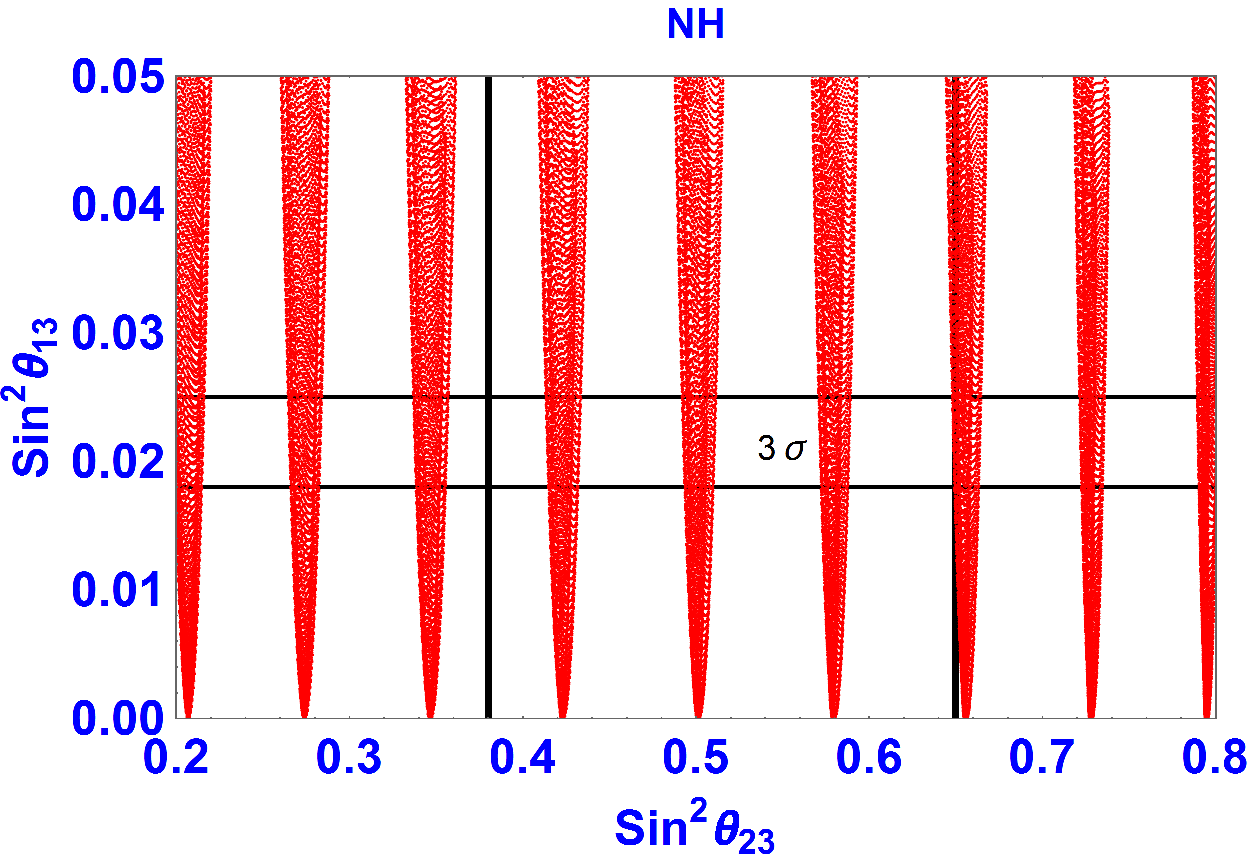} \\
\includegraphics[width=0.5\textwidth]{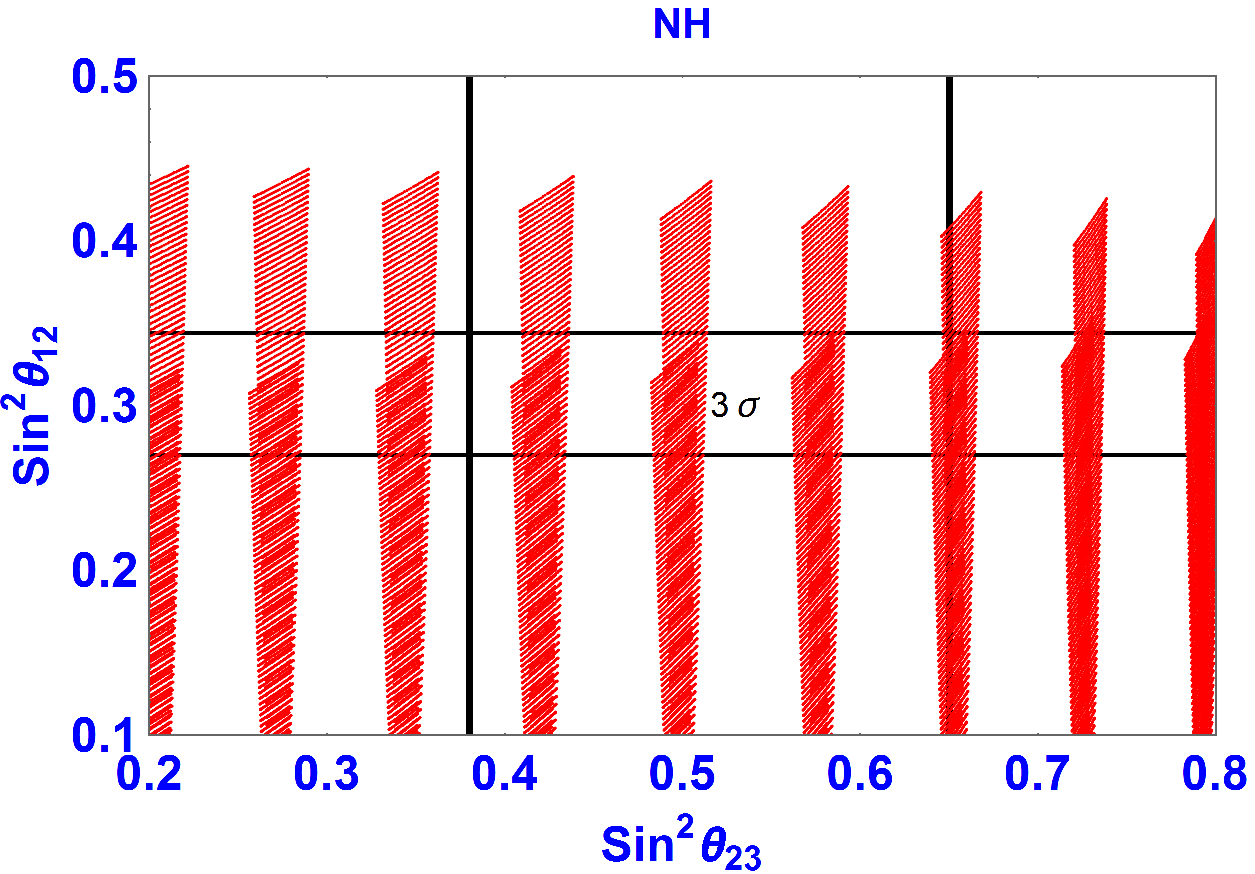} &
\includegraphics[width=0.5\textwidth]{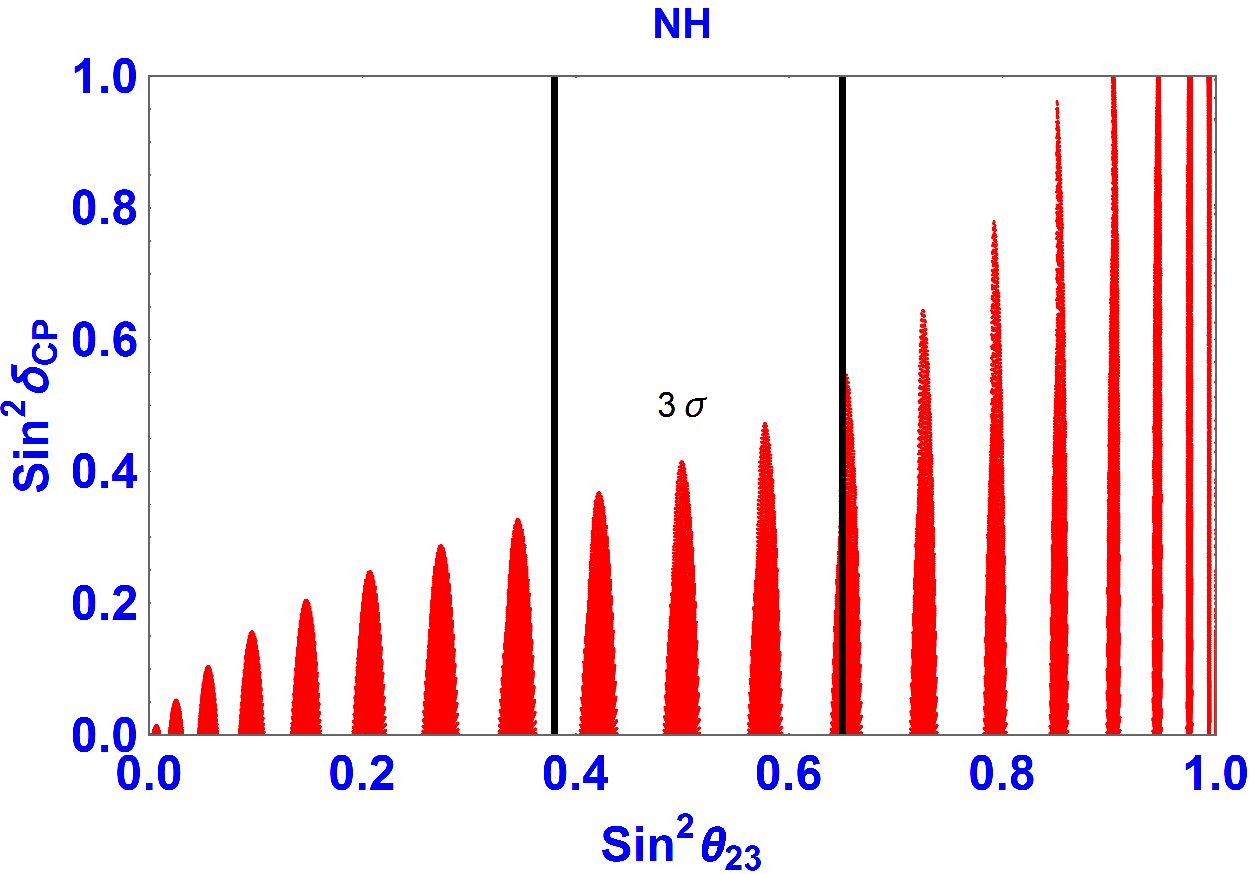}
\end{array}$
\end{center}
\caption{Correlation between the mixing angles $\sin^2{\theta_{12}}, \sin^2{\theta_{13}}, \sin^2{\theta_{23}}$ and Dirac CP phase $\sin^2{\delta_{CP}}$ in scenario II}
\label{fig4}
\end{figure}
We find that for inverted hierarchy, all the mixing angles can not be generated simultaneously within their $3\sigma$ allowed ranges. For normal hierarchy this is possible. We show the mixing angles $\theta_{12}, \theta_{13}, \theta_{23}$ and $\delta_{CP}$ as a function of the lightest neutrino mass $m_1$ in figure \ref{fig2}. It can be seen that the correct values of $\theta_{12}, \theta_{13}$ can be generated simultaneously only when $m_1 \sim 0.002$ eV. The mixing angle $\theta_{23}$ and Dirac CP phase $\delta_{CP}$ can however, be generated in the correct $3\sigma$ range for all values of $m_1$. We also show the variation of the mixing angles and CP phase as a function of $\theta = \theta^{\nu}_{23}-\theta^l_{23}$ in figure \ref{fig3}. It can be seen that the angles $\theta_{12}, \theta_{13}$ are not sensitive to the value of $\theta$ and can be generated in the correct $3\sigma$ ranges for any value of $\theta$. On the other hand, the mixing angle $\theta_{23}$ remains in the $3\sigma$ range only when $\theta \sim 0.75-1.0$ and $\theta \sim 2.25-2.5$. The Dirac CP phase is sensitive to the value of $\theta$ with $\sin^2{\delta_{CP}} \sim 1$ when $\theta \sim \pi/2$ and falling down to zero as $\theta \rightarrow 0, \pi$. It can be seen that the constraint on $\theta$ from the point of view of keeping $\theta_{23}$ within the correct range forces $\sin^2{\delta_{CP}}$ to approximately lie in the range $(0.35-0.50)$. This becaomes more clear from the $\sin^2{\delta_{CP}}-\sin^2{\theta_{23}}$ plot shown in figure \ref{fig4}. Figure \ref{fig4} also shows the relative correlation between $\theta_{13}-\theta_{12}, \theta_{13}-\theta_{23}, \theta_{12}-\theta_{23}$. It can be seen that the scenario under study narrowly allows simultaneous generation of mixing angles $\theta_{13}$ and $\theta_{12}$ in their correct $3\sigma$ ranges. We also check the variation of the mixing angles and Dirac CP phase with the parameter $\delta$ in the mixing matrix and found no correlation between them.

We can also construct the lepton mass matrices using the formalism discussed above. Assuming the charged lepton mass matrix to be Hermitian we can write it down as 
$$ M_l = U_l M^d_l U^{\dagger}_l $$
where $U_l = R^{l}_{23}R^{l}_{13}R^{l}_{12}$ and $M^d_l = \text{diag}(m_e, m_{\mu}, m_{\tau})$. Thus, the charged lepton mass matrix contains only two free parameters, the angle $\theta^l_{23}$ and the CP phase $\delta^l$. Similarly, the neutrino mass matrix can also be written as 
$$ M_{\nu} = U_{\nu} M^d_{\nu} U^{\dagger}_{\nu} $$
where $U_{\nu} = R^{\nu}_{23}R^{\nu}_{13}R^{\nu}_{12}$ and $M^d_{\nu} = \text{diag}(m_1, m_2, m_3)$. It should be noted that for symmetric Majorana neutrino mass matrices, the mass matrix can be constructed as
$$ M_{\nu} = U_{\nu} M^d_{\nu} U^T_{\nu} $$
However, there are additional Majorana CP phases apart from the Dirac CP phase in such a scenario. We do not pursue such a scenario in this work. The neutrino mass matrix therefore contains three free parameters, the lightest neutrino mass, the angle $\theta^{\nu}_{23}$ and the CP phase $\delta^{\nu}$. It is interesting to note that, although $m_1$, $\theta = \theta^{\nu}_{23}-\theta^l_{23}$ and $\delta = \delta^l = \delta^{\nu}$ are restricted to specific range of values in order to generate correct leptonic mixing angles, one still has the freedom to choose either $\theta^{\nu}_{23}$ or $\theta^l_{23}$ (but not both) arbitrarily. From the plots shown in figure \ref{fig3}, we can see that $\theta = \theta^{\nu}_{23}-\theta^l_{23} \approx \pi/4$ is a valid point consistent with neutrino data. Let us consider a particular combination of $(\theta^{\nu}_{23}, \theta^l_{23})$ to realize this which simpilifies our calculation. Let $(\theta^{\nu}_{23}, \theta^l_{23}) = (\pi/2, \pi/4)$. Also consider the lightest neutrino mass to be $0.002$ eV which is also consistent with the requirement of producing correct leptonic mixing angles. This allows us to calculate the other two neutrino masses using the best fit values of neutrino mass squared differences \cite{schwetz14}. These values are calculated to be $m_2 = 0.0088$ eV, $m_3 = 0.049$ eV for normal hierarchy. In terms of the Wolfenstein's parameter $\lambda$, the neutrino mass ratios obey
$$ \frac{m_1}{m_2} \approx \lambda, \;\; \frac{m_1}{m_3} \approx \frac{4}{5}\lambda^2 $$
The charged lepton mass hierarchies can also be expressed in terms of $\lambda$ as 
$$ \frac{m_e}{m_{\mu}} \approx 2 \lambda^4, \;\; \frac{m_{\mu}}{m_{\tau}} \approx \frac{6}{5}\lambda^2 $$
Using the values $(\theta^{\nu}_{23}, \theta^l_{23}) = (\pi/2, \pi/4)$, we can now construct the charged lepton as well as neutrino mass matrix in terms of Wolfenstein parameter $\lambda$. They are given by
\begin{equation}
M_l=\left(
\begin{array}{ccc}
 \frac{32 \lambda ^6}{5} & e^{i\delta}\lambda ^3+\frac{6 \lambda ^4}{5}+ \mathcal{O}(\lambda^8) & \lambda ^3-\frac{6 \lambda ^4}{5}e^{-i\delta}+ \mathcal{O}(\lambda^8) \\
 e^{-i\delta}\lambda ^3+\frac{6 \lambda ^4}{5}+ \mathcal{O}(\lambda^8) & \frac{1}{10} \left(5+6 \lambda ^2+ \mathcal{O}(\lambda^7)\right) & \frac{e^{-i\delta}}{10} \left(5-6 \lambda ^2+ \mathcal{O}(\lambda^{10})\right) \\
 \lambda ^3-\frac{6 \lambda ^4}{5}e^{i\delta}+ \mathcal{O}(\lambda^8) & \frac{e^{i\delta}}{10} \left(5-6 \lambda ^2+ \mathcal{O}(\lambda^{10})\right)
& \frac{1}{10} \left(5+6 \lambda ^2+ \mathcal{O}(\lambda^7)\right) \\
\end{array}
\right)
\end{equation}
\begin{equation}
M_{\nu}=\left(
\begin{array}{ccc}
 \frac{1}{5} \lambda ^2 \left(12-4 \lambda +\lambda ^2\right) & -\frac{2 e^{i\delta}\lambda  \left(-5+8 \lambda ^2+ \mathcal{O}(\lambda^3)\right)}{5 \sqrt{5}}
& -\frac{2}{5} e^{-i\delta}\lambda ^{3/2} \left(2-3 \lambda +\lambda ^2\right) \\
 -\frac{2 e^{-i\delta}\lambda  \left(-5+8 \lambda ^2+ \mathcal{O}(\lambda^3)\right)}{5 \sqrt{5}} & \frac{1}{25} \left(25+ \mathcal{O}(\lambda^4)\right) & \frac{4 e^{-2i\delta}(2-3 \lambda +\lambda ^2 ) \lambda ^{5/2}}{5 \sqrt{5}} \\
 -\frac{2}{5} e^{i\delta}\lambda ^{3/2} \left(2-3 \lambda +\lambda ^2\right) & \frac{4 e^{2i\delta}(2-3 \lambda +\lambda ^2) \lambda ^{5/2}}{5 \sqrt{5}} & \frac{1}{5}
\lambda  \left(4-4 \lambda +5 \lambda ^2\right) \\
\end{array}
\right)
\end{equation}
These structures may indicate the presence of additional flavor symmetries like Frogatt-Nielsen type \cite{frogniel} which we do not investigate in this present work.

It is worth checking whether we still have some freedom left to adjust the free parameters in such a way to reduce the mass matrices into simpler form. In particular, we check whether the mass matrices can have zeros, so called texture zero mass matrices \cite{texturezero}. These texture zeros not only make the theory more predictive and minimal but could also indicate the presence of additional symmetries behind them. For illustrative purpose, we derive the conditions for some one-zero textures in the Majorana neutrino mass matrix in this work. 
\begin{itemize}
\item $$(M_{\nu})_{11} = \frac{m_1 \left(m_1^2+3 m_2 m_3-m_1 \left(2 m_2+m_3\right)\right)}{m_2 m_3} $$
$(M_{\nu})_{11}=0$ is possible if $\lvert m_1 \rvert \approx 0.057, 0.009, 0.0$ eV. However these values are outside the range of $m_1$ required to produce correct leptonic mixing angle as seen from figure \ref{fig2}. Thus this particular one-zero texture is not possible in this scenario.

\item 
$$(M_{\nu})_{12} = \frac{1}{m_2}\sqrt{1-\frac{m_1}{m_3}} (-c^{\nu }{}_{23}
\left(m_1-m_2\right) m_2 \sqrt{\frac{m_1 \left(-m_1+m_2\right)}{m_2^2}} $$
$$ +e^{i \delta } \sqrt{\frac{m_1}{m_3}} \left(m_1^2-2 m_1 m_2+m_2 m_3\right)
s^{\nu }{}_{23}) $$
The constraint $(M_{\nu})_{12}=0$ gives rise to two equations involving three free parameters $m_1, \delta, \theta^{\nu}_{23}$. Since neutrino parameters are not sensitive to the choice of $\delta$ and we also have a freedom in choosing $\theta^{\nu}_{23}$ provided $\theta = \theta^{\nu}_{23}-\theta^l_{23}$ lies in the desired range as discussed above, this particular one-zero texture can be realized in these frameworks.
\end{itemize}
Similar conclusions can be drawn for other possible one-zero textures as well. Since we have three free parameters in the neutrino mass matrix, it will be worth investigating whether two-zero textures are possible in this framework. Similarly we can also study the possibility of simpler charged lepton mass matrix with texture zeros and constrain the mixing angle $\theta^l_{23}$. These we leave for future investigations.

\section{Conclusion}
\label{conclude}
We have studied the possibility of relating lepton mixing angles to lepton mass hierarchies. Guided by the existence of such relations in the quark sector, we also consider the mixing angles associated with charged lepton mass matrix and neutrino mass matrix to be related to the corresponding mass ratios. We show that, if we assume three mixing angles each in charged lepton and neutrino sectors to be related to three respective mass ratios, then the PMNS leptonic mixing matrix in the absence of CP violation can be constructed with one free parameter, the lightest neutrino mass. We vary this lightest neutrino mass both for normal and inverted hierarchical neutrino masses and show that all the three leptonic mixing angles can not be generated simultaneously within their $3\sigma$ range. We check that this overall conclusion does not change even if we take the leptonic Dirac CP phase into account. We then consider the validity of the relation between mixing angles and mass hierarchies to be restricted to only two mixing angles in each sector, charged lepton and neutrino. For simplicity, we consider 1-2 and 1-3 mixing angles to be related to the corresponding mass ratios and 2-3 mixing angles $\theta^l_{23}, \theta^{\nu}_{23}$ are kept as a free parameters. We also consider the origin of leptonic CP violation in the 2-3 sector and insert a phase term $\delta$ in the 2-3 rotation matrices. We then calculate the leptonic mixing angles and Dirac CP phase as a function of $\theta = \theta^{\nu}_{23}-\theta^{l}_{23}, \delta$ and the lightest neutrino mass. We show that for normal hierarchical neutrino masses, we can generate all the three leptonic mixing angles simultaneously in their correct $3\sigma$ range provided the lightest neutrino mass $m_1$ approximately lie around $0.002$ eV. We have also constrained the leptonic Dirac CP phase to the range $\sin^2{\delta_{CP}}\sim(0.35-0.50)$ so as to generate all the neutrino mixing angles in the correct $3\sigma$ range. Instead of taking 2-3 mixing angles as free parameters, one could take either 1-2 or 1-3 mixing angles as free parameters. In that case, the leptonic mixing angles will be functions of three free parameters instead of two in the minimal scenario discussed here. 

We then construct the charged lepton mass matrix with two free parameters, the $2-3$ mixing angle $\theta^l_{23}$ and the phase $\delta$ assuming it to be symmetric. We also construct the neutrino mass matrix upto three free parameters, the lightest neutrino mass, the 2-3 mixing angle $\theta^{\nu}_{23}$ and the phase $\delta$. To arrive at a simpler structure of mass matrices, we then assume specific values of $(\theta^{\nu}_{23}, \theta^l_{23}) = (\pi/2, \pi/4)$ and lightest neutrino mass $m_1=0.002$ eV consistent with correct PMNS mixing angles from the plot shown in figure \ref{fig2}, and derive the mass mass matrices in terms of Wolfenstein parameter $\lambda$. 

We further make a few remarks on the possibility of realizing some specific texture zeros in the Majorana neutrino mass matrix. We show that $(M_{\nu})_{11}=0$ texture is not possible whereas $(M_{\nu})_{12}=0$ is allowed within this framework. Similar studies can also be done to see the possibility of two-zero textures in the neutrino mass matrix. One can also constrain $\theta^l_{23}$ so as to allow texture zeros in the charged lepton mass matrix. It should be noted that we have considered the violation of the relation between mass ratios and mixing angles and origin of leptonic CP phase to be only in the 2-3 sector, whereas they can occur in 1-2 or 1-3 sector as well. We leave such a detailed study considering 1-2 or 1-3 mixing angles as free parameters instead of just $2-3$ mixing angles, possibility of texture zeros in both charged lepton and neutrino mass matrices to future work. These textures not only make the framework more predictive and minimal but also hint at an underlying symmetry behind them. Future measurement of neutrino mixing angles with more precision, smaller $3\sigma$ ranges and determination of absolute neutrino mass should be able to falsify the scenarios discussed in this work. Discovery of inverted hierarchy in neutrino experiments will also rule out the scenario discussed in the present work. Similarly, experimental measurement of leptonic Dirac CP phase will also falsify our framework as it predicts a narrow range of $\delta_{CP}$ such that $\sin^2{\delta_{CP}}\sim(0.35-0.50)$. It is worth mentioning that this range is already disfavored by the recent hint of $\delta_{CP} = -\pi/2$ \cite{diracphase}. However, more data are required to confirm such a measurement.

\end{document}